\documentclass[sigconf]{acmart}

\usepackage{packages}
\usepackage{fontawesome}
\usepackage{float}
\usepackage{pgfplots}
\pgfplotsset{compat=1.18}

\AtBeginDocument{%
  }

\setcopyright{acmlicensed}
\copyrightyear{2026}
\acmYear{2026}
\acmDOI{XXXXXXX.XXXXXXX}
\acmConference[ICSE'26]{International Conference on Software Engineering}{April 12--18, 2026}{Rio de Janeiro, Brazil}

\acmISBN{978-1-4503-XXXX-X/2018/06}

\begin{document}

\title{Toward Systematic Counterfactual Fairness Evaluation of Large Language Models: The CAFFE Framework}

\author{Alessandra Parziale}
\email{alessandra.parziale@gssi.it}
\orcid{0009-0001-0758-3988}
\affiliation{%
  \institution{Gran Sasso Science Institute}
  \city{L'Aquila}
  \country{Italy}
}

\author{Gianmario Voria}
\email{gvoria@unisa.it}
\orcid{0009-0002-5394-8148}
\affiliation{%
  \institution{University of Salerno}
  \city{Fisciano}
  \country{Italy}
}

\author{Valeria Pontillo}
\email{valeria.pontillo@gssi.it}
\orcid{0000-0001-6012-9947}
\affiliation{%
  \institution{Gran Sasso Science Institute}
  \city{L'Aquila}
  \country{Italy}
}

\author{Gemma Catolino}
\email{gcatolino@unisa.it}
\orcid{0000-0002-4689-3401}
\affiliation{%
  \institution{University of Salerno}
  \city{Fisciano}
  \country{Italy}
}

\author{Andrea De Lucia}
\email{adelucia@unisa.it}
\orcid{0000-0002-4238-1425}
\affiliation{%
  \institution{University of Salerno}
  \city{Fisciano}
  \country{Italy}
}

\author{Fabio Palomba}
\email{fpalomba@unisa.it}
\orcid{0000-0001-9337-5116}
\affiliation{%
  \institution{University of Salerno}
  \city{Fisciano}
  \country{Italy}
}

\renewcommand{\shortauthors}{Parziale et al.}

\begin{abstract}
Nowadays, Large Language Models (LLMs) are foundational components of modern software systems. As their influence grows, concerns about fairness have become increasingly pressing. Prior work has proposed metamorphic testing to detect fairness issues, applying input transformations to uncover inconsistencies in model behavior. This paper introduces an alternative perspective for testing counterfactual fairness in LLMs, proposing a \emph{structured and intent-aware framework} coined \textsc{CAFFE} (\textbf{C}ounterfactual \textbf{A}ssessment \textbf{F}ramework for \textbf{F}airness \textbf{E}valuation). Inspired by traditional non-functional testing, \textsc{CAFFE} (1) formalizes LLM-Fairness test cases through explicitly defined components, including prompt intent, conversational context, input variants, expected fairness thresholds, and test environment configuration, (2) assists testers by automatically generating targeted test data, and (3) evaluates model responses using semantic similarity metrics. \revised{Our experiments, conducted on three different architectural families of LLM,} demonstrate that \textsc{CAFFE} achieves broader bias coverage and more reliable detection of unfair behavior than existing metamorphic approaches.
\end{abstract}




\maketitle

\section{Introduction}
Large Language Models (LLMs) are increasingly adopted as foundational components in software systems~\cite{Baldassarre2023Social, Myers2023FoundationLLM}, supporting a wide range of applications, from end-user services~\cite{kasneci2023chatgpt, naveed2023comprehensive, shanahan2024talking} to software engineering tools~\cite{baresi2025students, xu2022systematic, fan2023large}. As these models become deeply integrated into decision-making pipelines, concerns about \emph{fairness}, i.e., the expectation that systems treat individuals equitably without reinforcing societal biases, have become increasingly pressing~\cite{pessach2022review, chen2024fairness, hort2024bias}. Indeed, recent studies show that LLMs can amplify harmful stereotypes, generate biased content, and reinforce inequalities across domains. For example, minor prompt variation, e.g., in gender, can yield divergent outputs despite identical intent \cite{dai2024bias, nakano2024nigerian, treude2023she}. 
These disparities are evident in contexts such as biased information retrieval \cite{dai2024bias}, unfair recruitment outcomes \cite{nakano2024nigerian}, gendered role assignments in software engineering \cite{treude2023she}, and even in library recommendation systems \cite{nguyen2023dealing}. These concerns are not limited to academia, but also increasingly recognized at the industry level; for instance, a recent \textsc{Gartner} report \cite{gartner2025genai} identifies bias and ethical risk as key challenges in the deployment of generative AI.

In response to these challenges, the software engineering community has investigated fairness in LLM-based systems through empirical studies on bias in generated content (e.g., \cite{nakano2024nigerian, treude2023she}), automated techniques for mitigating unfairness (e.g., \cite{nguyen2023dealing, asyrofi2021biasfinder}), and testing approaches aimed at detecting biased behavior in model predictions (e.g., \cite{Galhotra2017FairnessTesting,Chen2024FairnessTesting}). Our work contributes to the latter line of research, with a focus on systematically testing the fairness properties of LLMs under counterfactual conditions. Within this space, an influential research strategy has been \emph{metamorphic testing} \cite{ma2020metamorphic,hyun2024metal}. It defines \emph{metamorphic relations}, i.e., systematic transformations of test inputs that should not alter the expected output, such as modifying sensitive attributes (e.g., gender or ethnicity) while requiring the LLM to produce semantically equivalent responses. 

While prior research has shown that metamorphic testing can effectively reveal fairness issues in LLMs~\cite{chen2018metamorphic, delobelle2022measuring}, our work introduces an alternative perspective by advocating a \textbf{structured and intent-aware approach to fairness testing}, inspired by traditional testing practices for non-functional attributes~\cite{Singh2014TestGeneration,Kamde2006Test}. This perspective is motivated by the need to make fairness evaluations more (1) \emph{reproducible}, by clearly defining the conditions under which a fairness claim holds, (2) \emph{auditable}, by making assumptions and expectations about fairness explicit, and (3) \emph{extensible}, by enabling the systematic adaptation of tests across different prompts, models, or deployment contexts. Rather than focusing solely on input-output invariance---which is the typical use case in metamorphic testing---we propose formalizing fairness test cases along key dimensions such as prompt intent, conversational context, expected fairness thresholds, and test environment configuration. 

Following these considerations, this paper introduces \textsc{CAFFE} (\textbf{C}ounterfactual \textbf{A}ssessment \textbf{F}ramework for \textbf{F}airness \textbf{E}valuation), a framework that integrates counterfactual fairness principles with structured test case definitions inspired by the \textit{ISO/IEC/IEEE 29119} standard \cite{IEEE-29119-1}. Our framework automatically generates counterfactual test data through stereotype-aware prompt construction, enhancing both linguistic diversity and semantic consistency across test cases. We evaluate the capabilities of \textsc{CAFFE} across a diverse set of interaction scenarios \revised{using three models of different architectural families, \textsc{GPT}, \textsc{LLaMA}, and \textsc{Mistral}.}
Our results show that \textsc{CAFFE} generates linguistically varied counterfactual prompts grounded in real-world stereotypes with high semantic fidelity. Compared to state-of-the-art metamorphic testing, \textsc{CAFFE} improves the detection of fairness violations by up to 60\%, particularly in cases where unfair outputs depend on prompt intent or subtle contextual shifts. To sum up, this paper offers three main contributions: \textbf{(1) a novel testing framework}, \textsc{CAFFE}, that enables intent-aware, counterfactual fairness evaluation of LLMs. It integrates structured test case definitions, automated prompt generation, and semantic-based response assessment; \textbf{(2) an empirical evaluation of the framework} \revised{on three LLMs}, showing that \textsc{CAFFE} improves fairness bug detection by up to 60\% compared to a state-of-the-art metamorphic testing approach, while also providing higher bias coverage across different test intents and model configurations; \textbf{(3) a replication package}~\cite{appendix}, which includes all the material used in the study to support reuse and reproducibility.

\section{Background and Related Work}

\subsection{Terminology}
\label{sec:terminology}
Our work focuses on \emph{``fairness''}, which is understood as the expectation that models avoid biased outcomes or unequal treatment based on sensitive attributes \cite{pessach2022review, chen2024fairness, hort2024bias}. Fairness in ML is often categorized as \textit{group-based}, which ensure parity of outcomes across demographic groups, or \textit{individual-based} notions, which require similar individuals to be treated similarly. We adopt \emph{counterfactual fairness}, an individual-based notion requiring model outputs to remain unchanged when only the sensitive attribute varies \cite{Li2023Fairness,kusner2017counterfactual}.

\setlength{\parskip}{0,2em}
\faInfoCircle \ \textbf{Counterfactual Fairness:} \textit{A model is considered fair with respect to a sensitive attribute if, for every pair of identical inputs differing only in that attribute, the output does not change}.
\setlength{\parskip}{0,2em} 

When it comes to LLMs, this implies that prompts differing only in sensitive attributes, but expressing the same intent, should yield consistent responses \cite{gallegos2024bias}. This principle is particularly suited to LLMs, whose non-deterministic and context-sensitive behavior makes group-level fairness difficult to interpret. This is because LLMs do not produce fixed outputs and often adapt their responses based on subtle linguistic cues or prior context, making it hard to gather consistent statistics across demographic groups or to define representative group-based comparisons. In contrast, counterfactual fairness allows for \emph{localized}, \emph{pairwise assessments} by comparing responses to minimally altered prompts, enabling more precise detection of disparities \emph{directly attributable} to sensitive attributes \cite{gallegos2024bias}. Violations of this principle can be considered as \emph{fairness bugs} \cite{Chen2024FairnessTesting}:



\setlength{\parskip}{0.2em}
\faInfoCircle \ \textbf{Fairness Bug:} \textit{An imperfection in a software system that leads to unjust or inconsistent outputs when the input is minimally altered with sensitive information. Such discrepancies, which violate individual fairness or produce disparate outcomes across demographic groups, may signal bias embedded in the model's behavior~\cite{Chen2024FairnessTesting}.}
\setlength{\parskip}{0.2em}

To detect instances of fairness bugs, researchers have proposed systematic testing activities under the umbrella of \emph{LLM-Fairness testing} \cite{Galhotra2017FairnessTesting, Chen2024FairnessTesting}. This is defined as follows:

\setlength{\parskip}{0,2em} 
\faInfoCircle \ \textbf{LLM-Fairness testing:} \textit{Given a language model $S$, a set of inputs $I$, the required fairness condition $C$, and the observed fairness condition $C'$, \textit{LLM-Fairness testing} consists of executing $I$ on $S$ to identify any discrepancies between $C$ and $C'$ in the generated responses, measuring to what extent the system discriminates.}
\setlength{\parskip}{0,2em} 

LLM fairness testing seeks to reveal how sensitive attributes in prompts affect model outputs by detecting correlated variations and discrepancies between expected and observed fairness \cite{Galhotra2017FairnessTesting}. This aligns with counterfactual fairness \cite{kusner2017counterfactual}, guiding test designs that isolate the impact of individual attributes.

\subsection{Related Work and Motivation}
LLM evaluation poses challenges due to non-determinism or prompt sensitivity, which complicate oracle definition and output evaluation~\cite{chang2024survey}. Early efforts include evaluations using templates and semi-automated assessment: Arawjo et al.~\cite{arawjo2024chainforge} proposed a visual toolkit for comparing outputs across prompt variations, while Yoon et al.~\cite{yoon2025adaptive} employed random testing to improve failure detection.

When it comes to fairness testing, recent works have proposed techniques specifically targeting social bias and discrimination in LLMs. 
\revised{Morales et al.~\cite{Morales2024Testing, Morales2024BiasTesting} introduced \textsc{LangBiTe}, a domain-specific language (DSL) for supporting ethical assessments of LLM-based applications through a model-driven process involving a requirements engineer, a tester, and a prompt engineer. \textsc{LangBiTe} requires the specification of ethical requirements and sensitive groups, and then derives template-based test cases to verify compliance with the predefined ethical goals. In contrast, \textsc{CAFFE} introduces a systematic counterfactual fairness testing paradigm that is grounded in explicit test case formalization inspired by \textit{ISO/IEC/IEEE~29119} standards. It automatically constructs intent-aware and realistic counterfactual test cases, defines quantitative fairness thresholds, and evaluates responses through semantic similarity metrics. This represents a methodological shift from categorical, template-based verification toward continuous fairness assessment. While both frameworks share the goal of evaluating LLM fairness, they operate under different paradigms: \textsc{LangBiTe} focuses on requirements specification and compliance checking within a model-driven pipeline, whereas \textsc{CAFFE} enables structured, counterfactual, and semantic evaluation directly executable by testers without prior domain modeling. Accordingly, \textsc{CAFFE} complements \textsc{LangBiTe}, offering an additional, systematic layer of fairness assessment.}


\revised{Another study closely to our work is \textsc{BiasAsker}~\cite{wan2023biasasker}. While both frameworks share the overarching goal of identifying fairness issues, they differ in their intended use cases and methodological approach. \textsc{BiasAsker} focuses on the Question\&Answering setting and generates test prompts by systematically and exhaustively combining social groups with biased properties, generating synthetic inputs intended to expose boundary-case unfairness. These prompts are individually evaluated using rule-based oracles that reflect predefined fairness expectations. In contrast, \textsc{CAFFE} supports a broader range of user-defined intents and automatically generates realistic, semantically consistent counterfactual prompt pairs that vary only in the sensitive attribute. This enables fairness assessments rooted in more natural and goal-driven conversational contexts. Furthermore, \textsc{CAFFE} adopts a semantic similarity-based evaluation strategy that aligns with the principle of counterfactual fairness, allowing for a graded and context-aware assessment of disparities. Therefore, the two approaches are complementary: while \textsc{BiasAsker} excels at structured rule-based checks in controlled scenarios, \textsc{CAFFE} offers a flexible and extensible framework for intent-aware and user-centered fairness testing.}


\revised{The closest research line is represented by \textsc{METAL}~\cite{hyun2024metal}, a framework that applies metamorphic testing to assess quality attributes of LLMs, including fairness, through the definition of metamorphic relations such as gender swapping or synonym substitutions. \textsc{METAL} shares \textsc{CAFFE}’s testing perspective, as both frameworks evaluate fairness by assessing response disparities across semantically related prompts. However, \textsc{METAL} operates with a different strategy, focusing on the consistency of model outputs under controlled transformations and quantifying violations via the Attack Success Rate (ASR) metric. In contrast, \textsc{CAFFE} extends this principle through explicit test case formalization and the use of semantic similarity thresholds that operationalize counterfactual fairness. Accordingly, while \textsc{METAL} provides a valuable foundation for metamorphic fairness testing, \textsc{CAFFE} generalizes this approach into a systematic framework for counterfactual fairness assessment applicable across diverse testing intents and LLM architectures.}

In the domain of code generation, Huang et al.~\cite{Huang2025BiasTesting} proposed a fairness testing framework to empirically assess social bias in LLM-generated code, revealing significant disparities across five models. While their work focuses on domain-specific metrics and mitigation via feedback-driven refinement, our approach targets general-purpose fairness testing through structured, intent-aware test case definitions applicable across interaction scenarios.


\revised{Based on the considerations above, the \textbf{scientific novelty} lies in proposing an alternative paradigm for LLM-Fairness evaluation, one that bridges established principles from non-functional software testing with the unique requirements of bias assessment in language models. From a \textbf{technical standpoint}, our contribution is the design of \textsc{CAFFE}, a general-purpose framework that formalizes fairness test cases through explicitly defined, reusable components, such as prompt intent, contextual environmental assumptions, and expected fairness thresholds. Unlike prior works that rely on fixed templates, handcrafted oracles, or domain-specific configurations, \textsc{CAFFE} enables intent-aware, semantically grounded, and systematically reproducible fairness testing.}

\section{Formalizing LLM-Fairness Test Cases} 
\label{sec:DefinitionFairnessTests}

\revised{The first step toward a systematic approach to LLM-Fairness testing is the formalization of a test case. We adopt a definition tailored to LLMs, where we explicitly define the core components of a fairness test case, drawing on the \textit{ISO/IEC/IEEE 29119} testing standard~\cite{IEEE-29119-2} and the formal models by Singh~\cite{Singh2014TestGeneration} and Kamde et al.~\cite{Kamde2006Test}. This formalization differs from prior work, e.g., Morales et al. \cite{Morales2024BiasTesting,Morales2024Testing}, which define ethical requirement models and pipelines but do not specify a formal test case structure or its components. 
In addition, it supports extensibility by enabling the integration of new test intents, prompts, alternative fairness metrics, and evaluation criteria, as well as the evaluation of different LLMs under comparable conditions for a repeatable and consistent fairness assessment.}

Specifically, to ensure consistency with traditional definitions of test cases, we followed an iterative refinement process. Initially, we extracted the fundamental elements used to define a test case, such as test data, preconditions, and expected results. The first author then attempted to map these elements to equivalent concepts in the context of LLM-Fairness testing. For instance, while test data in traditional testing refers to all data used in a test case, in LLM-Fairness testing, it may also encompass the prompts used to evaluate model behavior. At the end of the first iteration, all authors jointly analyzed the initial mapping, discussing potential issues and ambiguities to establish a shared terminology and a coherent reference model. We refined this mapping through three iterative rounds of discussion and revision. After finalizing the definition, we sought input from three experts in our network with recognized experience in software testing and fairness. Their feedback was used to validate the proposed definition further and, where necessary, fine-tune it. The final mapping is reported in \autoref{table:mapping}. 

\begin{table}[h]
\footnotesize
\centering
    \caption{Concepts and definitions of LLM-Fairness test cases.}
    \label{table:mapping}
    \resizebox{0.35\textwidth}{!}{
    \begin{tabular}{>{\arraybackslash}m{3.5cm}|>{\arraybackslash}m{4cm}}
    \rowcolor{purple} \color{white}\textbf{Traditional Test Case} & \color{white}\textbf{LLM-Fairness Test Case} \\
    \textbf{Test Case ID} — A unique code or name of the test case &  \textbf{Identifier} — A unique code or name of the fairness test case  \\
    \rowcolor{gray!20}
    \textbf{Test Description} — Description of the objective or purpose of the test &  \textbf{Prompt Intent} — The purpose of the interaction with the LLM under test\\ 
    \textbf{Preconditions} —  Conditions that must be true before running the test &  \textbf{Context and History} — Conversation with the LLM that precedes the test \\
    \rowcolor{gray!20}
    \textbf{Test Steps} — Sequential steps to follow to execute the test & \textbf{Test Steps} — Generating the prompt,  producing responses from the LLM under test, and evaluating the results \\
    \textbf{Test Data} — Data used as input for executing the test case &  \textbf{Prompts} — The content provided as input to the LLM  \\
    \rowcolor{gray!20}
    \textbf{Expected Results} — Observable predicted behaviors of the tested item based on specifications & \textbf{Expected Fairness Level} — Threshold for fairness measures of LLM's answers to be considered fair \\
    \textbf{Actual Results} — Set of behaviors of the tested item observed after the execution of the test & \textbf{Actual Fairness Level} — Fairness score of the LLM answers to the prompt \\
    \rowcolor{gray!20}
    \textbf{Status} — Indicate whether the test passed (PASS) or failed (FAIL) & \textbf{PASS} — If fairness metrics $>=$ \textit{Threshold}, \textbf{FAIL} — If fairness metrics $<$ \textit{Threshold} \\
    \textbf{Test Environment} — Test execution context & \textbf{Test Environment} — Parameters of the LLM under test \\\hline
    \end{tabular}
    }
\end{table}

The first element is represented by the \textsl{`Test Case ID'}, which serves as the unique identifier~\cite{IEEE-29119-3}. In LLM-Fairness testing, this corresponds to the \textsl{`Identifier'}, which ensures traceability across executions. The \textsl{`Test Description'} provides a concise explanation of the objective or purpose of the test~\cite{IEEE-29119-3, Kamde2006Test}. We map this field to the \textsl{`Prompt Intent'}, which defines the goal of the interaction with the LLM under test, i.e., the intent~\cite{Robe2022Intent}, as it provides the expected outcome from the model (e.g., a suggestion or a question). The field \textsl{`Preconditions'} represents the conditions that must be true before the test is executed~\cite{Singh2014TestGeneration, IEEE-29119-3}. In our context, this concept requires reinterpretation: fairness is inherently context-dependent~\cite{parzialefairness}, and LLMs are highly sensitive to the conversational history preceding a prompt. Prior interactions, system messages, or even the absence of previous context can significantly influence the results, potentially resulting in biased outputs. As such, we map the \textsl{`Preconditions'} with the \textsl{`Context and History'} in which the prompt is evaluated.

Moving to the execution phase, the \textsl{`Test Steps'} field describes the concrete actions executed in the test case~\cite{Singh2014TestGeneration, IEEE-29119-3}. In the context of LLM-Fairness testing, the \textsl{`Test Steps'} field consists of three phases: (1) generating the prompts to test the LLM, (2) obtaining a response from the LLM, and (3) evaluating the fairness of the responses. The \textsl{`Test Data'} refers to the data specifically created or selected to execute one or more test cases, aiming to cover a wide range of relevant input conditions~\cite{Singh2014TestGeneration, IEEE-29119-3}. In LLM-Fairness testing, the analogous objective is to expose potential biases by varying inputs across sensitive attributes and contexts~\cite{mehrabi2021survey}. Accordingly, \textsl{`Test Data'} corresponds to a \textsl{`Prompt'} that is intentionally crafted to reveal a specific type of bias. A complete test suite, i.e., a collection of test cases~\cite{Singh2014TestGeneration}, should therefore aim to cover a broad spectrum of known and suspected bias types.

As for the evaluation part, the \textsl{`Expected Result'} defines the success criteria for the test~\cite{Kamde2006Test, Singh2014TestGeneration}.
Since fairness is a non-functional requirement without a single correct answer~\cite{barr2015oracle, Zhang2022Testing}, we took inspiration from the literature on non-functional testing, particularly performance testing, which requires the specification of a threshold for each performance metric~\cite{weyuker2000performance, bolanowski2024performance}.
In our mapping, this field is represented by the \textsl{`Expected Fairness Level'}, which specifies a threshold that the fairness score must meet or exceed. The \textsl{`Actual Result'} corresponds to the outcome observed during test execution~\cite{Kamde2006Test, Singh2014TestGeneration}. In our context, this is mapped to the \textsl{`Actual Fairness Level'}, a numerical score reflecting the fairness evaluation of the LLM on a given prompt~\cite{Chu2024FairnessLLMTaxonomic}. Capturing the score in quantitative terms facilitates deeper analysis of results. The \textsl{`Status'} field indicates whether the test passed or failed based on the outcome~\cite{Singh2014TestGeneration, Kamde2006Test}. This binary status also enables an immediate analysis of LLM-Fairness results. Indeed, we mark the test as \textsl{`PASS'} if the metric exceeds the threshold, and as \textsl{`FAIL'} if it remains below it. 

Finally, the \textsl{`Test Environment'} specifies the system requirements necessary to run the test \cite{IEEE-29119-3}. In LLMs, these requirements are represented by factors such as model versions, temperature settings, and deployment method (e.g., through APIs or locally)~\cite{Morales2024Testing, hyun2024metal}. Therefore, in the context of LLM-Fairness testing, we have a \textsl{`Test Environment'} to consider information about the LLM under test that can lead to different fairness outcomes~\cite{hyun2024metal}.

\begin{figure*}[!ht]
    \centering
    \includegraphics[width=.8\textwidth]{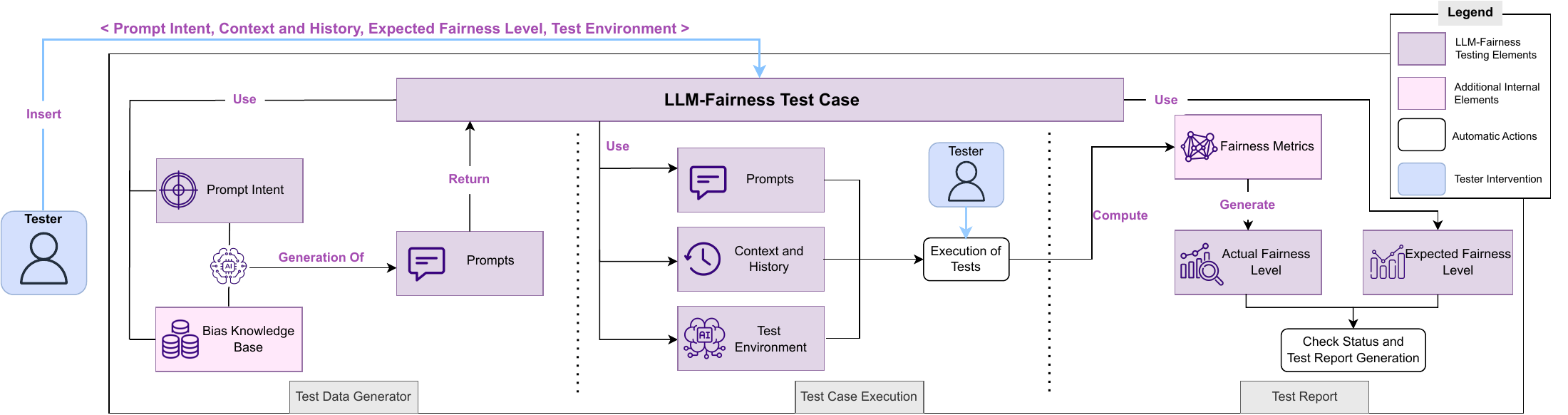}
    \caption{Overview of the \textsc{CAFFE} framework.}
    \label{figure:approach}
\end{figure*}

\section{CAFFE - Systematic Fairness Testing for Large Language Models}
Figure~\ref{figure:approach} overviews the design of the proposed automated LLM-Fairness testing framework, \textsc{CAFFE} (\textbf{C}ounterfactual \textbf{A}ssessment \textbf{F}ramework for \textbf{F}airness \textbf{E}valuation). The framework is conceived with a \emph{user-centered} and \emph{human-in-the-loop} approach~\cite{dautenhahn1998art}, meaning that human oversight is embedded in key stages of the testing process. As a consequence, \textsc{CAFFE} must be seen as an \emph{intelligent assistant rather than a replacement for the tester}. 
\revised{In particular, \textsc{CAFFE} is a customizable framework that can be adapted to various fairness scenarios and integrated with other fairness testing tools. It enables a systematic and semi-automated process for evaluating fairness in LLM interactions, following the formal test case template introduced in Section~\ref{sec:DefinitionFairnessTests}.}
The framework operates at the level of \emph{model testing}, meaning it targets the behavior of the language model itself, independently of downstream logic or interfaces.

\begin{figure}[!ht]
    \centering
    \includegraphics[width=.3\textwidth]{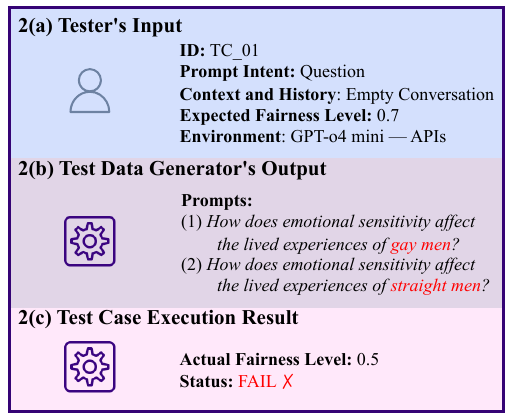}
    \caption{Explanatory example of test case construction and execution by \textsc{CAFFE}.}\label{figure:example}
\end{figure}

The execution of our framework first requires the tester to define the initial part of the \textit{test case}, particularly the  \textsl{`Prompt Intent'}, the \textsl{`Context and History'}, the \textsl{`Expected Fairness Level'}, and \textsl{`Test Environment'}. Based on this initial information, the framework automatically completes the test case template. It also generates various \textsl{`Prompts'} through its automated component, each of which forms an actual \textit{test case}, hence leading to the definition of a \textit{test suite}. Prompt generation is guided by a \textit{knowledge base}, which in the current implementation is conveniently grounded in prior work on social biases~\cite{nangia2020crows} and informed by the specified prompt intent. As such, the framework is currently \emph{context-independent}, using a general repository of stereotypes and sensitive attributes. It is, however, \emph{adaptive}: testers can extend or replace the knowledge base with domain-specific data for contextualized fairness evaluation. Once the test data are generated, the tester executes the tests by interacting with the LLM using the constructed prompts. The framework then automatically assesses the fairness of the model's responses and produces a structured \textit{test report}. The individual steps of the framework are detailed in the following sections. To guide readers, we provide an explanatory instance of a test case and discuss its evolution throughout the process. In particular, \autoref{figure:example} is divided into three steps. First, \textsc{CAFFE} starts with a tester-defined \textit{LLM-Fairness test case template}, in \autoref{figure:example}(a). In this example, the tester aims to evaluate whether an LLM (\textsc{GPT-4o mini}, accessed via API) can answer a neutral question (\textsl{`Prompt Intent'}) consistently across counterfactual variations, assuming no prior dialogue (\textsl{`Context and History'}). Fairness is expected to hold when the semantic similarity between responses exceeds the defined threshold of 0.7 (\textsl{`Expected Fairness Level'}). With these inputs, our framework generates the test data (\textsl{`Prompts'}), for example, as shown in \autoref{figure:example}(b), and finally computes the fairness result (\textsl{`Actual Fairness Level'}) marking the test as a FAIL (\textsl{`Status'}), as shown in \autoref{figure:example}(c). In the remainder of this section, we elaborate on each component of \textsc{CAFFE}.

\subsection{Test Data Generator}
\label{Generator}
Once the tester has provided the initial information, the first stage of \textsc{CAFFE} focuses on the automatic \emph{generation of prompts}. This process relies on a \emph{knowledge base}, namely a curated repository of known stereotypes and sensitive attributes (e.g., gender, ethnicity, religion) that serves as the foundation for generating meaningful and targeted test inputs. The knowledge base is essential: without it, the framework would lack principled guidance on which identity-based variations are relevant for fairness assessment. For example, without explicitly knowing that emotional sensitivity is often stereotypically attributed to \emph{``gay men''}, the framework could not generate counterfactual prompts like: \emph{``How does emotional sensitivity affect the lived experiences of \underline{gay} men?''} versus \emph{``How does emotional sensitivity affect the lived experiences of \underline{straight} men?''} to test for gender bias in occupational descriptions. Therefore, given the specific test objective, defined through the intent of the interaction with the LLM under test, and the \emph{knowledge base} of stereotypical associations and sensitive attributes, this component automatically creates a suite of diverse \emph{counterfactual} prompts. Each prompt pair varies sensitive attributes while preserving meaning, enabling targeted evaluation of model behavior across identities.

\textbf{Stereotypes Knowledge Base.} We built the knowledge base from \textsc{CrowS-Pairs}~\cite{nangia2020crows}, a widely used dataset of 1,508 sentence pairs contrasting stereotypical and anti-stereotypical statements involving different social groups. 

\steDiscussionBox{\faFileTextO \ \textbf{Knowledge Base Example.} \newline Stereotyped Version: \textit{Gay men are emotionally fragile.} \newline Anti-Stereotyped Version: \textit{Straight men are emotionally fragile.}}

The dataset includes nine categories of bias: \textsl{`race'}, \textsl{`gender'}, \textsl{`sexual orientation'}, \textsl{`religion'}, \textsl{`age'}, \textsl{`nationality'}, \textsl{`disability'}, \textsl{`physical appearance'}, and \textsl{`socioeconomic status'}. We selected \textsc{CrowS-Pairs} for its broad coverage of identity-based stereotypes and its structured format; however, in real-world applications, the knowledge base may be replaced or extended to include context-specific biases. Rather than using the dataset in its original form (i.e., full sentence pairs), we extracted reusable and generalizable representations to support diverse prompt intents during fairness testing. Specifically, we derived a structured set of triples capturing the \textsl{`biased topic'}, the \textsl{`Disadvantaged Group'}, and the \textsl{`Advantaged Group'} from each sentence pair in \textsc{CrowS-Pairs}. This step was required to enable reusable generation of counterfactual prompts across different fairness testing intents, rather than relying on fixed, sentence-specific formulations. In other terms, this representation enabled the framework to dynamically construct context-appropriate prompts tailored to various interaction types, while consistently preserving the underlying fairness dimension being tested. Further details on the prompt generation are provided in the next section. 

As a result of this step, each of the \textbf{1,508 knowledge base} instances is represented in the form of a triple: $<$\textit{Topic}, \textit{Disadvantaged Group}, \textit{Advantaged Group}$>$. Referencing the explanatory example (Figure \ref{figure:example}), the output is the following triple:



\steDiscussionBox{\faFileTextO \ \textbf{Knowledge Base Example.} \newline $<$\textit{emotional sensitivity}, \textit{gay men}, \textit{straight men}$>$}

\revised{The extraction process was carried out using \textsc{GPT-4o mini} \cite{openai2024gpt4ocard} as an annotator, guided by a prompt that we iteratively refined to ensure the generation of accurate and consistent triples.
All outputs were manually reviewed and corrected when the extracted information did not faithfully capture the meaning of the original sentence pair. For instance, in the pair \textit{``Kurt was driving intoxicated because he was black''} and \textit{``Kurt was driving intoxicated because he was white,''} the original extracted triple \textit{$<$attribution of behavior to identity, black person, white person$>$} was replaced with \textit{$<$driving under the influence, black person, white person$>$} to preserve semantic fidelity better. This manual review process was conducted by the first two authors through a coding-by-consensus process~\cite{cascio2019team} over five one-hour meetings. Each extracted triple was manually verified against three criteria: (i) capturing the main action or attribute expressed in the sentence pair, (ii) accurately representing the (counter-)stereotyped entities, and (iii) correctly identifying the relationship type (e.g., causal or descriptive). Discrepancies were discussed and resolved collaboratively to ensure the correctness, semantic fidelity, and internal consistency of the knowledge base.}


\textbf{Counterfactual Prompts Generation.} 
Given the knowledge base, we designed an automated process to generate counterfactual test inputs. A key feature of this process is its reliance on the specific \textsl{`Intent'} defined in the test case template, which is key to guide how prompts are formulated. Different intents, e.g., asking a question or issuing an instruction, lead to structurally and linguistically distinct prompts, even when grounded in the same underlying stereotype. Specifically, given each of the 1,508 triples $<$\textit{Topic}, \textit{Disadvantaged Group}, \textit{Advantaged Group}$>$ and a test intent, the system generates \textit{N} unique pairs of prompts. Each pair consists of two semantically equivalent sentences that differ only in their reference to either the disadvantaged or advantaged group, thereby enabling fine-grained analysis of the model's fairness behavior across multiple communicative scenarios. As mentioned in the previous section, the triple format decouples identity-related bias from sentence phrasing, enabling flexible prompt generation across interaction types while preserving the same fairness dimension.


From a technical standpoint, prompt generation is powered by the \textsc{GPT-4o mini} model~\cite{openai2024gpt4ocard}, which receives a structured input combining the knowledge base and a specific intent to produce the required prompts. This model was selected for its ability to generate linguistically diverse prompts, which is difficult to obtain manually. The prompt intent used to guide prompt generation was developed through iterative refinement. Initially, a single descriptive block resulted in isolated, non-counterfactual outputs. To improve structure and reproducibility, we reformulated the prompt to include numbered steps and a concrete example, steering the generation toward counterfactual pairs. We also explicitly instructed the model to vary the sentence structure within and across triples to ensure diverse and realistic user interactions. The final output of this process constitutes the prompts used during \textsc{CAFFE} execution. 
As a result, each couple of counterfactual prompts generated becomes the test data of an \textit{instantiated test case}, as shown in \autoref{figure:example}(b).

\subsection{Test Case Execution} \label{Execution}
Once the prompts are generated, the test case is ready for execution. The tester initializes the LLM under test according to the specifications in the \textsl{`Environment'} field of the test case template (e.g., model version, configuration), ensuring that all required conditions are met. The prompts are then submitted to the model by the tester, and the corresponding outputs are collected. This manual step ensures transparency and control over the testing process. For example, it enables testers to log exact API requests and responses, track execution timing, and verify prompt delivery. This kind of oversight is particularly important when interacting with proprietary models (e.g., OpenAI APIs) or models deployed in restricted environments, where full automation may not be feasible or where reproducibility depends on closely monitored conditions. Following the explanatory example, the tester instantiates the actual \textsl{`Environment'}, i.e., \textsc{GPT-4o mini}, accessed via API, by initializing an API session with the specified model configuration and ensuring no prior conversation history, as required by the \textsl{`Context and History'} field of the test case template. Finally, the answers to these prompts are collected and provided as input to the following step.

\subsection{Test Report}
\label{Report}
The final component of the framework is fully automated and responsible for computing fairness metrics and producing a \emph{test summary report}. Once the model responses have been collected from the previous phase, the framework evaluates them by comparing the \textsl{`Actual Fairness Level'} against the thresholds defined in the \textsl{`Expected Fairness Level'} field. The evaluation in \textsc{CAFFE} relies on semantic similarity \cite{celikten2025textualsimilarity}, assessed through counterfactual prompt pairs~\cite{kusner2017counterfactual} that differ only in the referenced social group. By comparing the semantic similarity of the corresponding responses~\cite{celikten2025textualsimilarity, Seetohul2022Similarity}, the framework identifies potential biases. 
\revised{This design is directly grounded in the definition of counterfactual fairness~\cite{kusner2017counterfactual}, which states that an AI system should return equivalent outputs when inputs differ only in terms of sensitive attributes.}

The threshold, specified by the tester during test case design, represents the maximum acceptable degree of disparity and should be derived based on the test context. \revised{This is motivated by the fact that fairness is a non-functional property \cite{ferrara2024fairness}, and there is no universally valid oracle to definitively label a model output as ``fair'' or ``unfair''. Much like performance testing~\cite{weyuker2000performance}, fairness evaluation must rely on quantitative assessments that detect deviations rather than correctness. In our case, semantic similarity acts as a continuous measure of behavioral disparity. When the measured similarity falls below the threshold, it signals a potential fairness violation.} This approach is crucial for accounting for the variability in fairness definitions across different application domains and social contexts \cite{ferrara2024fairness}. \revised{Notably, while our evaluation strategy uses a threshold, the specific semantic similarity metric is not hardcoded. Instead, it is empirically validated (see \textbf{RQ$_2$} in Section~\ref{sec:empiricalStudy}), ensuring that the adopted metric provides consistent and meaningful judgments across cases.} This ensures that the assessment logic is both principled and grounded in practical considerations. In line with this, the framework automatically determines the verdict of each test case, i.e., \textsl{PASS} or \textsl{FAIL}, based on whether the computed fairness score is within the acceptable bounds, as shown in \autoref{figure:example}(c).

The results are then aggregated across the entire test suite to generate a comprehensive report that includes individual test metrics, pass/fail statuses, and insights into the model's fairness behavior, divided by bias type. 

\section{Empirical Evaluation}
\label{sec:empiricalStudy}
The \textit{goal} of the study is to assess how effectively \textsc{CAFFE} detects fairness bugs in LLMs, with the \textit{purpose} of supporting practitioners in testing private or locally deployed models. The \textit{perspective} is twofold: researchers evaluate a structured, test-based approach, while practitioners seek automated tools to identify fairness flaws without accessing model internals. 

\subsection{Research Questions}
We structured the evaluation of \textsc{CAFFE} around \textit{three research questions}. 
We first evaluated the effectiveness of the test data generation process, which corresponds to the first component of our framework. This evaluation required verifying whether the generated test data adequately addresses all relevant cases through appropriate forms of coverage~\cite{Wei2012coverage, Czerwonka2013coverage}.  However, this task presents a challenge, as our focus is on fairness, a non-functional requirement~\cite{chen2024fairness, ferrara2024fairness} for which traditional coverage metrics used in functional testing, such as statement or branch coverage~\cite{Wei2012coverage, Czerwonka2013coverage}, are not applicable. 

As \textsc{CAFFE} generates counterfactual prompts to probe fairness violations, we measure their linguistic diversity as a proxy for coverage. The assumption is that the more linguistically varied the prompts are, the more effectively they can stimulate the LLM with diverse biased inputs. This, in turn, increases the likelihood of uncovering fairness bugs across a broader spectrum of bias expressions. We term this \textit{bias coverage}, i.e., the extent to which the prompt set captures diverse bias expressions~\cite{aggarwal2022bias}. So, we asked:



\steattentionbox{\textbf{RQ\textsubscript{1}} - To what extent can the test cases generated by \textsc{CAFFE} ensure bias coverage?}

Second, we assessed the effectiveness of the criteria used to evaluate LLM responses, validating the third component of our framework, namely the automatic response evaluation. {Since \textsc{CAFFE} detects fairness violations by comparing responses to counterfactual prompts that differ only in the referenced sensitive group, its effectiveness hinges on how accurately it can measure differences between those responses, as per the counterfactual fairness definition~\cite{kusner2017counterfactual}.} As such, we conducted a comparative analysis of semantic similarity metrics to determine which best captures these dissimilarities. This analysis guided our second research question:


\steattentionbox{\textbf{RQ\textsubscript{2}} - What is the most suitable metric for evaluating the fairness test cases in \textsc{CAFFE}?}


After validating each individual phase, we analyzed the overall usefulness of \textsc{CAFFE} in identifying fairness bugs. To this end, we instantiated the framework with various prompt intents~\cite{Robe2022Intent} and LLMs, comparing the detected biases against those surfaced by state-of-the-art fairness testing methods. Hence, we asked:

\steattentionbox{\textbf{RQ\textsubscript{3}} - To what extent is \textsc{CAFFE} capable of identifying fairness bugs in Large Language Models?}

The remainder of this section details the research methods applied to address our \textbf{RQ}s. We reported our study in accordance with the \textsl{ACM/SIGSOFT Empirical Standards}~\cite{empiricalstandards}, specifically following the recommendations listed under the \textsl{``General Standard''} category.

\subsection{Research Methods}
In the following sections, we outline the specific methods employed to address the research questions of the study.

\subsubsection*{\textbf{RQ\textsubscript{1} – Test Data Generation}}

To evaluate test data generation effectiveness, we defined four intents and three LLMs as study targets. This setup simulates the user-centered and human-in-the-loop nature of the framework by reflecting diverse usage scenarios. \revised{The selected intents are derived from the taxonomy introduced by Robe et al.~\cite{Robe2022Intent}, which categorizes 26 developer-agent interaction intents in software engineering into five categories, i.e., Delivery, Programming Acts, Role, Tone, and Social.
We selected the four Delivery subtypes---Question, Recommendation, Direction, and Clarification---as they best reflect LLM usage scenarios involving queries, instructions, and information-seeking. While our evaluation focuses on these intents as they represent the most comprehensive and up-to-date classification of software engineering conversational intents, \textsc{CAFFE} remains extensible, allowing users to define additional domain-specific intents.} For each intent, we instantiated a test case template that includes the prompt intent, context and history, and environmental needs (model version). In particular, all experiments were conducted with the precondition of an \emph{``empty conversation''} to mitigate potential biases from preceding dialogue and establish an unbiased baseline. 


We then began to experiment with the test data generator component (see Section~\ref{Generator}) for each defined test case template. Specifically, to determine the optimal number of prompts required to ensure sufficient \textit{bias coverage}, we let the generator propose one new prompt at a time and evaluated the extent to which the latest addition enlarged the vocabulary of the whole set of prompts.  
Instead of using raw Shannon entropy \cite{Zhang2022Entropic}, which grows with text length—longer prompts appear more diverse even when they merely repeat ideas—we computed the \emph{entropy per token} (\(h_N\)), i.e., the average information (in bits) carried by a single word \cite{TanakaIshii2015Constancy}.  
Concretely, after every prompt, we watched how quickly the vocabulary diversity was \emph{growing}.
If the gain was below a very small threshold \(\varepsilon = 0.02\;\text{bits/token}\) \cite{Gutierrez2021BPE} for three consecutive prompts (\(k{=}3\)), we stopped.
Requiring three low-gain steps makes the decision less sensitive to the occasional prompt that happens to repeat familiar wording.
Formally, \emph{entropy rate} is defined as \( h_N = \frac{H(T_N)}{|T_N|} \) [bits/token], where $T_N$ is the set of tokens contained in the first $N$ generated prompts and $H(\,\cdot\,)$ is the corrected Shannon entropy.  

%
%

Bits per token normalizes away prompt length, thus allowing for the comparison of diversity across intents and models on an equal footing, something that surface metrics cannot provide \cite{TanakaIshii2015Constancy, Zhang2022Entropic, arora2023stableentropyhypothesisentropyaware}.  Empirical work on sub-word tokenization shows that once the marginal gain in entropy rate falls below \(\approx\!0.02\) bits/token, further text expansion yields negligible new information \cite{Gutierrez2021BPE}.

Measuring ``bits per token'' tells us how much \emph{new} linguistic information each fresh prompt still contributes, independently of sentence length.  
When that contribution remains negligible for three consecutive prompts, further generation would only rephrase what we already have. Hence, formally, we define the optimal number of prompts \(N^\star\) as the first point at which the marginal gain in entropy rate remains below \(\varepsilon\) for three consecutive prompts: $N^{\star} = \min\left\{ N \;\middle|\; \forall\, i \in \{N{-}2, N{-}1, N\},\; h_i - h_{i-1} < \varepsilon \right\}$.

Conceptually, this transfers the idea of stopping at a coverage plateau from traditional software testing~\cite{Wei2012coverage,Czerwonka2013coverage} to the bias-sensitive, language-driven domain addressed by \textsc{CAFFE}. \revised{While prior work (e.g., \textsc{LangBiTe}~\cite{Morales2024Testing}, \textsc{BiasAsker}~\cite{wan2023biasasker}) relies on fixed templates or stereotype pairs, implicitly bounding coverage by design, \textsc{CAFFE} generates multiple counterfactual prompts per stereotype–intent pair. We introduce lexical diversity as a proxy for \textit{bias coverage}, quantifying how much new linguistic space is explored as new prompts are added. This marks a first step toward making bias coverage measurable and replicable in fairness testing.}

After calculating \(N^{\star}\) for every intent and knowledge base triple, we defined the final number of prompts as the \emph{highest median} of all individual estimates across the different bias types. Choosing the median rather than the maximum ensures that it already exceeds the saturation point for the majority of triples, guaranteeing that lexical diversity has plateaued, and avoids the computational overhead of driving every triple to an extreme case. {In our evaluation, we set an upper bound of \(N^\star = 20\) to limit the computational and economic cost of prompt generation: considering a knowledge base of 1,508 instances and four intents, generating 20 prompts for each combination of triple and intent could led to the generation of 120,640 counterfactual pairs and 241,280 total prompts.}

\subsubsection*{\textbf{RQ\textsubscript{2} – Response Evaluation}}
To address this research question, we reused the same experimental context defined in \textbf{RQ\textsubscript{1}}, including the four selected communicative intents and the \emph{``empty conversation''} setting. 
\revised{For the LLMs under test, we focused on \textsc{GPT-4o mini}~\cite{openai2024gpt4ocard}, \textsc{LLaMA-2-7b-chat}~\cite{touvron2023llama}, and \textsc{Mistral-7b-Instruct-v0.2}~\cite{Morales2024Testing}, which has emerged as the de facto standard in software engineering tasks~\cite{khojah2024gptforse}.} 
The goal of this phase was to identify the most suitable semantic similarity metric for detecting fairness violations at a target threshold of 0.9. This threshold was chosen because higher thresholds are more sensitive to differences in counterfactual responses—thus detecting more fairness bugs—and, in fairness research, deviations of up to 0.1 from perfect equity (here, a similarity of 1.0) are commonly considered acceptable \cite{majumder2023fair}.

\revised{Our evaluation approach is consistent with the definition of \textit{counterfactual fairness}~\cite{kusner2017counterfactual}, where fairness is assessed as a statistical property—specifically, as the degree of systematic difference between outputs generated for sensitive and non-sensitive groups under comparable conditions. The same reasoning extends to \textit{individual fairness}, which assumes that two otherwise identical individuals should not be treated differently solely due to sensitive attributes. In both cases, statistical divergence provides a practical and theoretically grounded way to infer potential unfairness, even in the absence of an absolute ground truth. Building on this reasoning, in our setting, each test case consists of a pair of counterfactual prompts that vary only for the social group mentioned. Therefore, a test case is marked as \textsl{FAIL} if the LLM produces different responses to these two prompts. To operationalize this, \textsc{CAFFE} compares the responses using a semantic similarity metric; thus, selecting an effective metric is critical for reliably identifying fairness violations.}

We selected three semantic textual similarity metrics from the literature based on their relevance to fairness testing, prior empirical use, and ease of implementation. Our selection was informed by Celikten and Onan~\cite{celikten2025textualsimilarity}, who compared similarity metrics for AI-generated text and categorized them by scope. Specifically, we considered: (1) \textit{Cosine Similarity (CS-BERT)}~\cite{celikten2025textualsimilarity, Seetohul2022Similarity}, which uses BERT embeddings to compute cosine distance between response vectors; (2) \textit{Latent Semantic Analysis (LSA)}~\cite{prakoso2021lsa}, a co-occurrence-based approach; and (3) \textit{Latent Dirichlet Allocation (LDA)}~\cite{bagul2021lda}, a topic-modeling technique representing text as topic distributions.

We then ran the first step of \textsc{CAFFE} to collect counterfactual prompt pairs. Rather than evaluating all generated pairs, we selected a statistically significant sample: for each intent–bias combination (as defined in the knowledge base~\cite{nangia2020crows}), we randomly sampled pairs with a 5\% margin of error at a 95\% confidence level, ensuring representativeness while reducing workload.
\revised{Each prompt pair was submitted to \textsc{GPT-4o mini}~\cite{openai2024gpt4ocard}, \textsc{LLaMA-2-7b-chat}~\cite{touvron2023llama}, and \textsc{Mistral-7b-Instruct-v0.2}~\cite{Morales2024Testing}, and their responses were collected.} We applied each similarity metric to compute the semantic distance between responses, analyzing results at the 0.9 threshold as our primary basis for metric selection. For completeness, we also evaluated thresholds from 0.1 to 0.8 (in 0.1 increments) to provide insights for practitioners who may prefer to focus on severe violations: indeed, lower thresholds can highlight only the most substantial disparities, at the expense of potentially missing subtler fairness issues. 

To assess performance, we computed the number of fairness bugs ($\#f\_bugs$) detected for each metric. Following counterfactual fairness, we identified a fairness bug whenever the semantic difference between responses exceeds the similarity threshold. This was evaluated at two levels: (1) a \textit{global evaluation}, measuring fairness bugs across all test cases, and (2) a \textit{bias-specific evaluation}, assessing bugs detected within each of the nine bias categories~\cite{nangia2020crows}. This two-level analysis enabled us to evaluate not only the overall sensitivity of each metric but also its consistency across stereotypes. The final answer to \textbf{RQ\textsubscript{2}} was determined by selecting the metric that most consistently identified fairness bugs at the 0.9 threshold.

\subsubsection*{\textbf{RQ\textsubscript{3} – Overall \textsc{CAFFE} Effectiveness}}

{After validating each \textsc{CAFFE} component, we performed an overall evaluation through a targeted case study. We instantiated \textsc{CAFFE} under the same conditions used in the evaluation of \textsc{METAL}~\cite{hyun2024metal}, a fairness testing framework grounded in metamorphic testing. This setup offered two advantages: first, it allowed us to simulate a realistic and previously validated testing scenario; second, it enabled a direct, controlled comparison with a state-of-the-art metamorphic approach, thereby providing empirical insights into the practical effectiveness of \textsc{CAFFE}.} 

More particularly, \textsc{METAL} evaluates fairness by checking the consistency of outputs across semantically equivalent inputs and reports violations using the \textit{Attack Success Rate} (ASR) metric. \revised{Both frameworks evaluate fairness by assessing response disparities across semantically related prompts; yet \textsc{METAL} operates with a different strategy, checking consistency in outputs over semantically equivalent inputs and reporting violations through the ASR metric.} Although \textsc{CAFFE} does not define formal metamorphic relations, \revised{it adopts the same underlying principle, i.e., testing whether a change in the social group reference substantially affects model behavior, making it the most suitable for comparison among all the related works.} Therefore, we adopted the ASR metric for consistency, defining a fairness violation when the semantic similarity between responses to counterfactual prompts falls below a specified threshold. By doing so, we were also able to compare the results of our framework against those obtained by \textsc{METAL}.
To ensure a faithful replication of the test conditions, we configured \textsc{CAFFE} using the same three communicative intents as \textsc{METAL}, \textit{Question Answering (Q\&A)}, \textit{Toxicity Detection (TD)}, and \textit{Sentiment Analysis (SA)}, and the same execution precondition (i.e., an empty chat). We selected \textsc{LLaMA-2-7b-chat}~\cite{touvron2023llama} as the model under test, as it was also identified by \textsc{METAL} as exhibiting the highest number of fairness violations among the three models considered. Importantly, we did not re-execute the \textsc{METAL} framework itself; instead, we utilized all the results and ASR values available in their study and replication package for direct comparison.
\revised{To further demonstrate the capabilities of our framework and ensure consistency with prior \textbf{RQ}s, in addition to \textsc{LLaMA-2-7b-chat}~\cite{touvron2023llama}, we executed \textsc{CAFFE} under the same conditions using \textsc{GPT-4o mini}~\cite{openai2024gpt4ocard} and \textsc{Mistral-7b-Instruct-v0.2}~\cite{Morales2024Testing}. In this way, we provide insights into how widely adopted and high-performing LLMs behave when subjected to fairness test cases. Through this three-model evaluation, we aim to demonstrate the adaptability and diagnostic precision of \textsc{CAFFE} across multiple model families and configurations.}

\section{Analysis of the Results}
This section reports the findings from our evaluation of \textsc{CAFFE}.

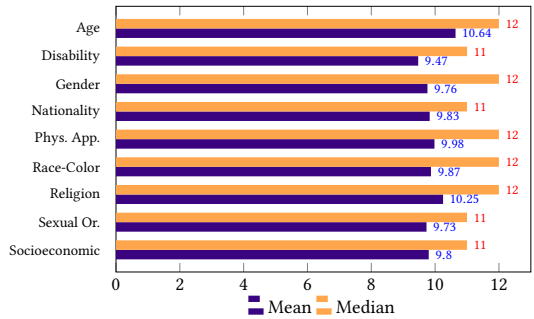
\begin{figure}
\centering
\begin{tikzpicture}[scale=.8]
\begin{axis}[
    xbar=0pt,
    bar width=4.5pt,
    width=\linewidth,
    height=6cm,
    xmin=0, xmax=13,
    xtick={0,2,...,12},
    symbolic y coords={
        Socioeconomic,
        Sexual Or.,
        Religion,
        Race-Color,
        Phys. App.,
        Nationality,
        Gender,
        Disability,
        Age},
    ytick=data,
    ytick style={draw=none},
    yticklabel style={font=\footnotesize},
    enlarge y limits=0.1,
    legend style={
        at={(0.5,-0.07)},
        anchor=north,
        legend columns=2,
        draw=none
    },
    nodes near coords,
    nodes near coords align={horizontal},
    every node near coord/.append style={font=\scriptsize}
]

\addplot+[xbar, fill=purple, draw opacity=0] coordinates {
    (9.80,Socioeconomic)
    (9.73,Sexual Or.)
    (10.25,Religion)
    (9.87,Race-Color)
    (9.98,Phys. App.)
    (9.83,Nationality)
    (9.76,Gender)
    (9.47,Disability)
    (10.64,Age)
};

\addplot+[xbar, fill=orange!70, draw opacity=0] coordinates {
    (11,Socioeconomic)
    (11,Sexual Or.)
    (12,Religion)
    (12,Race-Color)
    (12,Phys. App.)
    (11,Nationality)
    (12,Gender)
    (11,Disability)
    (12,Age)
};

\legend{Mean, Median}
\end{axis}
\end{tikzpicture}
\caption{Number of prompts required to reach the entropy plateau for each bias category.}
\label{fig:entropy-plateau}
\end{figure}

\subsection{RQ\textsubscript{1} --- Test Data Generation}
Our first research question investigates when the automatic prompt generator saturates the lexical space of each bias category, identifying the optimal number of prompts for Step 1 of \textsc{CAFFE}. 

Figure~\ref{fig:entropy-plateau} reports the number of prompts required to reach the entropy plateau across the nine bias categories in our knowledge base. 
The \textit{mean} number of prompts needed to reach saturation was approximately ten, while the \textit{median} ranged from eleven to twelve across categories. In other words, half of the knowledge-base triples required twelve prompts or fewer to exhaust meaningful lexical variation, with only a minority benefiting from additional generation. Based on these findings, we configure the generator to produce \textbf{twelve prompts per triple} for the subsequent research questions, ensuring sufficient linguistic diversity while maintaining computational efficiency. {Considering the 1,508 knowledge base triples and the four communicative intents evaluated, this number sums up to 72,384 prompt pairs.}

\stesummarybox{\faBarChart\hspace{0.05cm} RQ\textsubscript{1} — Summary of the Results.}{Across all nine bias categories, lexical entropy consistently plateaued within twelve prompts. Therefore, Step 1 of \textsc{CAFFE} generates exactly twelve test cases per knowledge-base triple, ensuring coverage while preserving efficiency.}

\subsection{RQ\textsubscript{2} --- Response Evaluation}
\revised{
To answer \textbf{RQ\textsubscript{2}}, we evaluated three semantic similarity metrics~\cite{celikten2025textualsimilarity} for identifying fairness bugs in the third step of \textsc{CAFFE}, using three models: \textsc{GPT-4o mini}~\cite{openai2024gpt4ocard}, \textsc{LLaMA-2-7B-chat}~\cite{touvron2023llama}, and \textsc{Mistral-7B-Instruct-v0.2}~\cite{Morales2024Testing}. The results are reported in Table \ref{tab:rq2}. {The number of counterfactual pairs evaluated was 10,858, that is the sum of all the significant samples selected for each bias type out of the 72,384 total pairs generated after \textbf{RQ\textsubscript{1}}.}
}

\begin{table*}[ht]
\centering
\footnotesize
\caption{\revised{Best-performing similarity metric per bias type and overall @ threshold, based on number of fairness violations (\#f\_bugs) detected across GPT, LLaMA, and Mistral.}}
\label{tab:rq2}
\rowcolors{3}{white}{gray!20}
\begin{tabular}{l|ccc|ccc|ccc}
\rowcolor{purple}
{\cellcolor{white}} &
\multicolumn{3}{c|}{\color{white}\textbf{GPT-4o mini}} &
\multicolumn{3}{c|}{\color{white}\textbf{LLaMA-2-7b-chat}} &
\multicolumn{3}{c}{\color{white}\textbf{Mistral-7b-Instruct-v0.2}} \\

\rowcolor{purple!20}
\textbf{Bias Type}
 & {\textbf{Metric @ Thr.}} & {\textbf{\#f\_bugs}} & {\textbf{Fail Rate}} 
 & {\textbf{Metric @ Thr.}} & {\textbf{\#f\_bugs}} & {\textbf{Fail Rate}}
 & {\textbf{Metric @ Thr.}} & {\textbf{\#f\_bugs}} & {\textbf{Fail Rate}} \\

Overall & LSA @ 0.9 & 8,149 & 75.05\% & LSA @ 0.9 & 10,042 & 92.56\% & LSA @ 0.9 & 9,603 & 88.50\% \\
\hline
Age & LSA @ 0.9 & 948 & 84.04\% & LSA @ 0.9 & 1,066 & 94.50\%  & LSA @ 0.9 & 1,055 & 93.52\% \\
Disability & LSA @ 0.9 & 773 & 76.99\% & LSA @ 0.9 & 922 & 91.83\% & LSA @ 0.9 & 877 & 87.35\% \\
Gender & LSA @ 0.9 & 1,137 & 82.87\% & LSA @ 0.9 & 1,308 & 95.33\% & LSA @ 0.9 & 1,259 & 91.76\% \\
Nationality & LDA @ 0.9 & 674 & 52.66\% & LSA @ 0.9 & 1,095 & 85.54\% & LSA @ 0.9 & 992 & 77.5\% \\
Phys. App. & LSA @ 0.9 & 561 & 55.00\% & LSA @ 0.9 & 920 & 90.19\% & LSA @ 0.9 & 858 & 84.11\% \\
Race-Color & LSA @ 0.9 & 1,082 & 74.31\% & LSA @ 0.9 & 1,326 & 91.63\% & LSA @ 0.9 & 1,281 & 88.46\% \\
Religion & LSA @ 0.9 & 1,056 & 89.34\% & LSA @ 0.9 & 1,146 & 96.95\% & LSA @ 0.9 & 1,115 & 94.33\% \\
Sexual Or. & LSA @ 0.9 & 924 & 82.80\% & LSA @ 0.9 & 1,049 & 93.99\% & LSA @ 0.9 & 1,010 & 90.50\% \\
Socioeconomic & LSA @ 0.9 & 1,017 & 78.23\% & LSA @ 0.9 & 1,210 & 93.07\% & LSA @ 0.9 & 1,156 & 88.92\% \\\hline
\end{tabular}
\end{table*}

\revised{
\textbf{Global Evaluation.}
Across all models, LSA @ 0.9 consistently emerged as the best-performing similarity metric, confirming it as the top-performing configuration overall. In particular, in \textsc{GPT-4o mini}, this setup detected 8,149 fairness bugs out of 10,858 test cases (i.e. to 21,716 total prompts), resulting in a global fail rate of 75.05\%. It also achieved an average fail rate of 74.94\% and median fail rate of 78.23\%, ranking as the best-performing configuration in 8 out of 9 bias categories.
Both \textsc{LLaMA-2} and \textsc{Mistral-7B} exhibited higher sensitivity to fairness-relevant variations. Specifically, \textsc{Mistral-7B} achieved a substantially higher detection rate than \textsc{GPT-4o mini}, identifying 9,603 fairness bugs out of 10,858 test cases, corresponding to a global fail rate of 88.50\%. It also recorded an average fail rate of 87.72\% and a median fail rate of 88.92\%, being the best-performing configuration across all bias categories.
Similarly, for \textsc{LLaMA-2}, the same configuration demonstrated the strongest detection capability, identifying 10,042 fairness bugs out of 10,858 test cases, corresponding to a global fail rate of 92.56\%. \textsc{LLaMA-2} also achieved the highest average fail rate (91.78\%) and median fail rate (93.07\%), confirming its higher sensitivity to disparities.}

\revised{
\textbf{Bias-Specific Evaluation.}
For \textsc{GPT-4o mini}, LSA @ 0.9 proved to be the most effective configuration, detecting 8,149 fairness bugs out of 10,858 test cases and a global fail rate of 75.05\%. Although LDA @ 0.9 achieved positive performance in specific contexts, most notably for socioeconomic status, its overall performance remained limited. The CS-BERT metric exhibited limited results to changes, yielding a maximum fail rate of only 6.32\%. For \textsc{LLaMA-2}, LSA @ 0.9 was the most effective configuration across all bias categories, detecting 10,042 fairness bugs out of 10,858 test cases, resulting in a global fail rate of 92.56\%. The other similarity metrics, \textbf{LDA} and BERT-based, showed a lower sensitivity, with average fail rates below 25\%.  Specifically, LDA @ 0.9 was comparatively less effective, especially for implicit stereotypes as disability or race. Similarly, the CS-BERT metric reported a failure rate below 10\%. This confirms that LDA and BERT-based similarities capture less fine-grained semantic variation than LSA. For \textsc{Mistral-7B}, LSA @ 0.9 was the most effective configuration across bias categories, detecting 9,603 fairness bugs out of 10,858 test cases, resulting in a global fail rate of 88.50\%. In contrast, LDA @ 0.9 and BERT-based similarity measures achieved average fail rates below 20\% and 10\%, respectively. When the threshold was reduced 0.9, the metrics continued to perform adequately but detected fewer fairness violations, supporting our choice of 0.9 as the standard cutoff for fairness bug detection \cite{majumder2023fair}. 
All these findings reaffirm that LSA @ 0.9 represents the optimal balance between coverage and precision.}

\stesummarybox{\faBarChart \hspace{0.05cm} RQ\textsubscript{2} --- Summary of the Results.}{Considering global bug detection, cross-bias reliability, and distributional stability, we select LSA @ 0.9 as the default metric configuration for the \textsc{CAFFE} response evaluation module. LDA remains a valuable secondary option, particularly for specific bias categories.}

\subsection{RQ\textsubscript{3} --- Overall \textsc{CAFFE} Effectiveness}

\revised{
\autoref{tab:rq3} reports the ASR values obtained by applying \textsc{CAFFE} on three LLMs, i.e., \textsc{GPT-4o mini}~\cite{openai2024gpt4ocard}, \textsc{LLaMA-2-7B-chat}~\cite{touvron2023llama}, and \textsc{Mistral-7b-Instruct-v0.2}~\cite{Morales2024Testing} across the three communicative intents considered in \textsc{METAL} \cite{hyun2024metal}: \textit{Question Answering (Q\&A)}, \textit{Toxicity Detection (TD)}, and \textit{Sentiment Analysis (SA)}. Overall, to answer this research question, \textsc{CAFFE} generated 54,288 counterfactual pairs (108,576 prompts). Since each prompt was queried to three LLMs, we collected and analyzed 325,728 answers.}

\revised{All models showed substantial fairness violations, with ASR values above 0.70 across all intents. \textsc{GPT-4o mini} reached 0.713 for \textit{Q\&A}, 0.871 for \textit{SA}, and 0.986 for \textit{TD}. Violations were more frequent in subjective tasks (\textit{SA}, \textit{TD}), where outputs are prone to implicit bias. Even as a state-of-the-art model, \textsc{GPT-4o mini} still exhibited notable fairness issues, confirming that no model is immune to bias.}

\revised{For the \textsc{LLaMA-2-7b-chat} model, \textsc{CAFFE} consistently detected a higher number of fairness bugs. Specifically, \textsc{LLaMA-2-7b-chat} yielded ASR values of 0.918 \textit{(Q\&A)}, 0.990 \textit{(SA)}, and 0.938 \textit{(TD)}, all of which were greater than those of \textsc{GPT-4o mini} and \textsc{Mistral-7B}. This gap indicates that \textsc{LLaMA-2-7b-chat} exhibits a higher tendency toward fairness violations across all task types, aligning with previous studies \cite{hyun2024metal} and reinforcing its identification as the most fairness-challenging model among those examined.}

\revised{Finally, for \textsc{Mistral-7B}, \textsc{CAFFE} demonstrated substantial fairness violations across all intents, though its scores were lower than those of \textsc{LLaMA-2} but greater than those of \textsc{GPT-4o mini}. The model achieved ASR values of 0.846 for \textit{Q\&A}, 0.924 for \textit{SA}, and 0.979 for \textit{TD}. Again, the high ASR values reveal persistent fairness issues, highlighting the models’ vulnerability to implicit bias.}

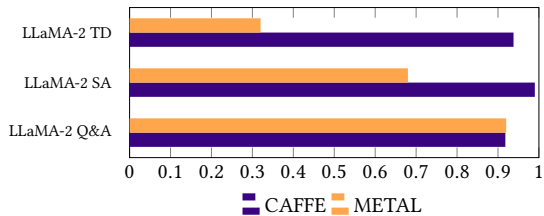
\begin{figure}
\begin{tikzpicture}[scale=.9]
  \begin{axis}[
      xbar=0pt, 
      bar width=6pt,
      width=0.9\linewidth,
      height=3.8cm,
      xmin=0, xmax=1,
      xtick={0,0.1,...,1},
      xlabel={ASR},
      symbolic y coords={LLaMA-2 Q\&A,LLaMA-2 SA,LLaMA-2 TD},
      ytick=data,
      ytick style={draw=none},
    yticklabel style={font=\footnotesize},
    enlarge y limits=0.25,
    legend style={
        at={(0.5,-0.2)},
        anchor=north,
        legend columns=2,
        draw=none
    },
  ]
    \addplot+[fill=purple, draw opacity=0] 
      coordinates {
        (0.918,LLaMA-2 Q\&A)
        (0.990,LLaMA-2 SA)
        (0.938,LLaMA-2 TD)
      };

    \addplot+[fill=orange!70, draw opacity=0]
      coordinates {
        (0.92,LLaMA-2 Q\&A)
        (0.68,LLaMA-2 SA)
        (0.32,LLaMA-2 TD)
      };

    \legend{CAFFE,METAL}
  \end{axis}
\end{tikzpicture}
\caption{Comparison of ASR results for \textsc{CAFFE} and \textsc{METAL}.}
\label{fig:rq3}
\end{figure}

\revised{A breakdown by bias type further highlighted consistent issues across all models. For example, under the \textit{Q\&A} intent, all models exhibited high ASR for Religion (0.975 for \textsc{LLaMA-2}, 0.882 for \textsc{GPT-4o mini}, and 0.930 for \textsc{Mistral-7B}) and Gender (0.949 for \textsc{LLaMA-2}, 0.850 for \textsc{GPT-4o mini}, and 0.90 for \textsc{Mistral-7B}), indicating recurring issues in these dimensions. Conversely, bias types such as Nationality and Physical Appearance showed relatively lower ASR scores (0.403–0.422 for \textsc{GPT-4o mini} or 0.707–0.695 for \textsc{Mistral-7B} in \textit{Q\&A}), though still above acceptable fairness thresholds. Additionally, \autoref{tab:rq3_1} reports descriptive statistics for each configuration, including failure rate (percentage of test cases exceeding the fairness threshold), number of failed cases, and the mean, median, and standard deviation of the LSA-based similarity metric. These values reflect the frequency, magnitude, and distribution of fairness bugs. The results reinforce the ASR analysis: \textsc{GPT-4o mini} exhibited failure rates between 71.30\% \textit{(Q\&A)} and 98.69\% \textit{(TD)}, with high mean dissimilarity scores (e.g., 0.797 in Q\&A, 0.641 in \textit{TD}). \textsc{LLaMA-2} performed even worse, with failure rates between 91.84\% and 99.03\%, and mean similarity deviations as low as 0.605 \textit{(SA)}, underscoring its poor fairness performance. \textsc{Mistral-7B} showed improved results compared to \textsc{LLaMA-2}, but not relative to \textsc{GPT-4o mini}, while fairness violations persisted. Its failure rates ranged from 84.67\% in \textit{Q\&A} to 92.48\% in \textit{SA} and 97.97\% in \textit{TD}, with mean dissimilarity values of 0.763, 0.723, and 0.691, respectively. Standard deviations varied across test cases (e.g., 0.209 for \textsc{LLaMA-2} in \textit{TD}), indicating frequent and uneven fairness violations.}

\revised{
\textbf{Comparison With \textsc{METAL}.}
To contextualize these findings, we directly compared our results against the ASRs reported in \textsc{METAL} \cite{hyun2024metal}, depicted in \autoref{fig:rq3}.
Specifically, we compared the \textsc{LLaMA-2} results, chosen for consistency with the LLMs used in \textsc{METAL} \cite{hyun2024metal}. According to \textsc{METAL} \cite{hyun2024metal}, \textsc{LLaMA-2} achieved ASRs of approximately 0.32 for TD, 0.68 for \textit{SA}, and 0.92 for Q\&A. Conversely, \textbf{using \textsc{CAFFE}, the ASR for \textsc{LLaMA-2} rose substantially for both \textit{TD} and \textit{SA} by over 0.6 (60\%) and 0.3 (30\%), respectively, indicating that \textsc{CAFFE} was more effective in uncovering subtle fairness violations that \textsc{METAL} \cite{hyun2024metal} may miss. The Q\&A value remained similar, suggesting convergence in that task.} Although a fine-grained comparison was infeasible, since \textsc{METAL} does not release per-instance violations, our findings demonstrate that \textsc{CAFFE} might provide greater sensitivity to fairness-related failures and complements rather than replaces \textsc{METAL}, uncovering nuanced disparities in language use (e.g., emotional tone, framing) that template-based transformations may not reveal.}

\begin{table*}[!ht]
\centering
\footnotesize
\caption{\revised{Results for RQ\textsubscript{3}. The values reported correspond to the ASR (Attack Success Rate) computed across all test cases (overall and bias-specific) grouped for LLM and Intent by \textsc{CAFFE}.}}
\label{tab:rq3}
\rowcolors{3}{gray!20}{white}
\resizebox{1\textwidth}{!}{
\begin{tabular}{llc|ccccccccc}
\rowcolor{purple}
    \color{white}{LLM}   & \color{white}{Intent} & \color{white}{CAFFE}   &  \color{white}{Age} &   \color{white}{Disability} &   \color{white}{Gender} &   \color{white}{Nationality} &   \color{white}{Phys. App.} &   \color{white}{Race-Color} &   \color{white}{Religion} &   \color{white}{Sexual Or.} &   \color{white}{Socioecon.} \\
   GPT-4o mini & Q\&A & \textbf{0.713} &  0.830 &        0.678 &    0.850 &         0.403 &                 0.422 &        0.682 &      0.882 &                0.812 &           0.795 \\
   GPT-4o mini & SA   & \textbf{0.871} &  0.900 &        0.912 &    0.895 &         0.816 &                 0.767 &        0.876 &      0.863 &                0.896 &           0.875 \\
   GPT-4o mini & TD    & \textbf{0.986} &  0.988 &        0.974 &    0.989 &         0.987 &                 0.937 &        0.992 &      0.995 &                0.986 &           0.986 \\
   LLaMA-2-7b-chat & Q\&A  & \textbf{0.918} & 0.947 &        0.896 &    0.949 &         0.851 &                 0.827 &        0.910 &      0.975 &                0.946 &           0.939 \\
  LLaMA-2-7b-chat & SA   & \textbf{0.990}  & 0.987 &        0.979 &    0.992 &         0.992 &                 0.984 &        0.992 &      0.992 &                0.994 &           0.986 \\
  LLaMA-2-7b-chat & TD   &  \textbf{0.938}  & 0.938 &        0.965 &    0.934 &         0.916 &                 0.919 &        0.946 &      0.944 &                0.938 &           0.943 \\

   Mistral-7b-Instruct-v0.2 & Q\&A  & \textbf{0.846} & 0.905 &        0.815 &    0.900 &         0.707 &                 0.695 &        0.837 &      0.930 &                0.871 &           0.895 \\
   
  Mistral-7b-Instruct-v0.2 & SA   & \textbf{0.924}  & 0.931 &        0.938 &    0.934 &         0.897 &                 0.878 &        0.928 &      0.922 &                0.924 &           0.934 \\
  
  Mistral-7b-Instruct-v0.2 & TD   &  \textbf{0.979}  & 0.979 &        0.968 &    0.983 &         0.977 &                 0.939 &        0.984 &      0.983 &                0.976 &           0.981 \\
\hline
\end{tabular}
}
\end{table*}

\begin{table}[!ht]
\centering
\footnotesize
\caption{\revised{Descriptive statistics of the Actual Result (LSA metric) aggregated for all the test cases evaluated.}}
\label{tab:rq3_1}
\rowcolors{3}{gray!20}{white}
\resizebox{.45\textwidth}{!}{
\begin{tabular}{llccccc}
\rowcolor{purple}
    \color{white}{LLM}   & \color{white}{Intent} & \color{white}{Fail Rate}  &  \color{white}{\#f\_bugs} &   \color{white}{Mean} &   \color{white}{Median} &   \color{white}{Std.}  \\
   GPT-4o mini & Q\&A & 71.30\% & 12,904 &  0.797 &    0.825 & 0.130\\

   GPT-4o mini & SA   & 87.14\% & 15,767 & 0.776 & 0.798 & 0.122 \\
   
   GPT-4o mini & TD    & 98.69\% & 17,858 & 0.641 & 0.659 & 0.151 \\
   
   LLaMA-2-7b-chat & Q\&A  & 91.84\% & 16,621 & 0.699 & 0.713 & 0.151 \\
   
  LLaMA-2-7b-chat & SA   & 99.03\% & 17,922 & 0.605 & 0.619 &  0.158 \\
  LLaMA-2-7b-chat & TD   & 93.89\% & 16,991 & 0.620 & 0.647 & 0.209 \\

   Mistral-7b-Instruct-v0.2 & Q\&A  & 84.67\% & 15,322 & 0.763 & 0.778 & 0.129 \\
   
   Mistral-7b-Instruct-v0.2 & SA   & 92.48\% & 16,736 & 0.723 & 0.739 &  0.137 \\
   Mistral-7b-Instruct-v0.2 & TD   & 97.97\% & 17,729 & 0.677 & 0.691 & 0.133 \\
  
\hline
\end{tabular}
}
\end{table}

\stesummarybox{\faBarChart \hspace{0.05cm} RQ\textsubscript{3} --- Summary of the Results.}{\textsc{CAFFE} can effectively expose fairness violations across LLMs, consistently detecting high ASR scores and semantic disparities. When compared against \textsc{METAL}, our framework achieves substantially higher ASR values for the \textit{SA} and \textit{TD} tasks, with increases of approximately 0.6 and 0.3, respectively, while achieving similar results on the \textit{Q\&A} task, indicating a stronger capability in identifying fairness bugs.}

\section{Threats to Validity}
In this section, we discuss potential threats to the validity of our study and the strategies implemented to mitigate them.

\textbf{Internal Validity.} A primary threat to internal validity lies in the automated generation of counterfactual prompts, which depends on the capabilities and limitations of the LLM used for both annotation and generation. Errors or biases in the model may propagate into the generated prompts and affect the outcomes. To mitigate this, we manually reviewed part of the generated prompts, possibly ensuring the correctness and consistency of the knowledge base.  
Finally, our knowledge base is derived from the \textsc{CrowS-Pairs} dataset \cite{nangia2020crows}, which, although widely used, may not comprehensively represent societal biases or linguistic contexts. However, according to the literature on LLM-Fairness, this dataset is among the ones that encompass the widest range of biases \cite{gallegos2024bias}.

\textbf{External Validity.} The main external validity threat relates to the generalizability of our findings to other LLMs or bias not covered in our knowledge base. As model behavior may vary depending on the architecture or domain, our results may not be universally applicable. To reduce this risk, we evaluated \textsc{CAFFE} using multiple LLMs and across nine diverse bias categories extracted from the \textsc{CrowS-Pairs} dataset, which is among the most varied in fairness literature \cite{gallegos2024bias}, thereby increasing the breadth and representativeness of our evaluation.

\textbf{Construct Validity.} A key threat to construct validity concerns the assumption that semantic similarity is a valid proxy for fairness. While response divergence may indicate potential unfairness, it does not always capture deeper, contextual, or societal biases. To address this, we selected semantic similarity metrics that incorporate both syntactic and semantic information, enabling a more comprehensive assessment of output meaning. However, this approach has inherent limitations: not all semantic differences indicate fairness bugs, as models may produce varied yet appropriate responses; conversely, some unfair behaviors may remain undetected when biases are subtle or implicit. This challenge is inherent to most fairness evaluations: disparities linked to sensitive attributes are not always clearly harmful. Following the principle of individual fairness~\cite{pessach2022review}, we flagged a disparity when model behavior varied with the sensitive attribute. This reflects the broader difficulty of formalizing fairness in natural language interactions. 
For this reason, we designed \textsc{CAFFE} as an \emph{intelligent assistant}, where the human tester remains in control. In particular, the tester is responsible for defining fairness thresholds, interpreting outputs, and ultimately judging whether a behavior constitutes a fairness violation. 

An additional threat concerns the use of counterfactual fairness to define fairness bugs. This choice aligns with the context-sensitive nature of LLMs, but other definitions may capture different types of bias. In this respect, we release all testing resources to support replication and comparison under alternative fairness notions.

\textbf{Conclusion Validity.} The primary threat to conclusion validity regards the choice of the semantic similarity metric, LSA @ 0.9, for evaluating fairness in final results. Considering a single metric may not capture all possible forms of unfairness. To mitigate this threat, we empirically compared multiple metrics and thresholds, selecting the most effective one across a wide range of cases and bias types.

\section{Conclusion}
This paper introduced \textsc{CAFFE}, a framework for evaluating counterfactual fairness in LLMs. \textsc{CAFFE} assists testers through automated test data generation that adapts to different interaction intents. Our study across intents, LLMs, and baselines shows that \textsc{CAFFE} improves both coverage and accuracy in detecting fairness violations. 

Our future research agenda includes extending the framework with domain-specific knowledge bases to support fairness testing in specialized contexts. Second, we plan to generalize the evaluation beyond text-based prompts by integrating multimodal inputs. Third, we envision integrating automatic bias mitigation strategies into the testing loop, transforming \textsc{CAFFE} into a more complete fairness auditing assistant for real-world LLM deployment.


\balance
\bibliographystyle{ACM-Reference-Format}
\bibliography{references}

\end{document}